\documentclass[twocolumn,showpacs,nofootinbib,10pt]{revtex4-1}

\usepackage{color,float,amssymb,amsmath,upgreek}
\usepackage{color}
\usepackage{graphicx}
\usepackage[dvipsnames]{xcolor}
\usepackage{graphicx}\usepackage[colorlinks,hyperindex]{hyperref}
\newcommand{\beq}{\begin{eqnarray}}
\newcommand{\eeq}{\end{eqnarray}}
\newcommand{\be}{\begin{equation}}
\newcommand{\ee}{\end{equation}}
\newcommand{\bea}{\begin{eqnarray}}
\newcommand{\eea}{\end{eqnarray}}

\newcommand{\ba}{\begin{eqnarray}}
\newcommand{\ea}{\end{eqnarray}}

\hypersetup{hidelinks,backref=true,pagebackref=true,hyperindex=true,colorlinks=true,breaklinks=true,urlcolor= blue}
\hypersetup{%
  colorlinks = true,
  linkcolor  = blue
}

\begin{document}

\title{Information-entropic analysis of Korteweg--de Vries solitons in the quark-gluon plasma}

\author{A. Goncalves da Silva}
\email{allan.goncalves@ufabc.edu.br}
 \affiliation{
Centro de Ci\^encias Naturais e Humanas, Universidade Federal do ABC - UFABC\\ 09210-580, Santo Andr\'e, Brazil.
}

\author{Rold\~ao~da~Rocha}
\email{roldao.rocha@ufabc.edu.br}
\affiliation{Centro de Matem\'atica, Computa\c c\~ao e Cogni\c c\~ao, Universidade Federal do ABC - UFABC\\ 09210-580, Santo Andr\'e, Brazil.}

%
%


\begin{abstract}

Solitary waves propagation of baryonic density perturbations, ruled by the Korteweg--de Vries equation in a mean-field quark-gluon plasma model, are investigated from the point of view of the theory of  information. A recently proposed  continuous logarithmic measure of information, called configurational entropy, is used to derive the soliton width, defining the pulse, for which the informational content of the soliton spatial profile is more compressed, in the Shannon's sense.

\end{abstract}



\maketitle

\section{Introduction}

Since the early decades of the past century, the information theory has been used in many areas of research. Even before the concept of information has arisen, the concept of entropy as a logarithmic measure of information had appeared in physics, representing at that time a breakthrough in the understanding of many-particle systems. The concept of informational entropy was developed with the work of Fisher \cite{fisher}, Wiener \cite{wiener} and Shannon \cite{shannon}, among others, between the middle of the 1920s and the end of the 1940s. Since then, it has been applied in many areas, from physics to computer science, mathematical statistics and communication theory. 
Besides all this success, the majority of the applications regarding the information entropy  use a discrete measure, instead of the continuous one that were proved to be a real threat. 
In fact, the so-called information dimension, that  underlies any distribution of data,  
yields an upper bound on the compression rate of any variable in a given distribution. The information dimension encodes the growth rate of the Shannon entropy the finer the space discretization is \cite{Bernardini:2016hvx}. 

In the last five years, a continuous logarithmic measure called configurational entropy (CE) has been used in the lattice approach as a new information tool, to study non-linear  physical systems \cite{Gleiser:2013mga,Gleiser:2014ipa,Sowinski:2015cfa,Gleiser:2012tu}. More precisely, the CE is a logarithmic measure, in the Fourier space, of the spatial complexity of spatially-bounded functions. It represents the exact measure of information that is necessary  to describe the spatial shape of functions with respect to their set of parameters. This approach has been used to study compact astrophysical systems 
 \cite{Gleiser:2013mga,Gleiser:2015rwa,Casadio:2016aum}, to derive the dominance of light-flavor mesons in dynamical AdS/QCD holographic models 
 \cite{Bernardini:2016hvx}, as well as to study scalar glueballs 
 \cite{Bernardini:2016qit}. Besides, the CE was employed to study AdS-Schwarzschild black holes, in a robust setup that corroborates to the  consistency of the Hawking-Page phase transition \cite{Braga:2016wzx}. Moreover, the (nuclear) CE was used in QCD models to establish the onset of the quantum regime in the color glass condensate \cite{Karapetyan:2016fai,Karapetyan:2017edu}. In addition, the informational entropy theory  
 was successfully used in particle physics to determine the Higgs boson mass \cite{Alves:2017ljt} and the axion mass in an effective theory at  low energy regimes \cite{Alves:2014ksa}. 
 
 Travelling solitary waves, or solitons  \cite{Correa:2015lla}, and other topological defects were already approached in the context of the configurational entropy 
 \cite{Correa:2015vka,Correa:2016pgr}.
Solitons were proposed to be studied as  solutions to non-linear equations of motion \cite{Brito:2001ga}. Although the analysis with respect to small perturbations was studied in other contexts \cite{Bazeia:1997zp,Bazeia:1999jt,dosSantos:2011zz}, in particular in the QCD \cite{Blas:2007dw}, among which other interesting applications have been found, we shall  focus on the configurational entropy approach. 

A special class of solitons is typified by the  Korteweg--de Vries (KdV) ones. KdV solitons
arise 
as a solution of a non-linear, dispersive partial differential equation, known as the KdV equation, whose  solutions are spatially localized pulses, of finite energy,  propagate with at most tiny distortions of their shape. The KdV equation has been frequently used to study waves in shallow-water and internal waves in oceans, to investigate acoustic waves propagating across crystal lattices and, more recently,  also to scrutinize waves in plasmas. In this context, it is useful to study the quark-gluon plasma (QGP), which  is a state of matter that is  produced either at extremely high temperatures, as in the early stage of the universe, or at very high density, as in the core of neutron stars. The QGP has been achieved in laboratory \cite{brokhaven}. 
Larger and longer living states in the QGP can consist of KdV solitons in the QCD plasma. Ref. \cite{Fowler} regarded perturbations on the baryons density in proton-nucleus collisions, where the incoming proton that might be absorbed by the nuclear fluid can generate a KdV soliton.
This setup was refined in Ref. \cite{fn5,fn4,fn6} with an equation of state ruling the nuclear matter, which is derived from a mean-field approximation to QCD in the case of zero-temperature and high density, called cold quark-gluon plasma (cQGP).
 The equations that govern the 1-dimensional relativistic fluid dynamics were studied in Ref. \cite{Fogaca:2011pk} in a context of perturbations beyond the linear regime, implying a  KdV equation. Hence, the CE entropic-information setup shall be used to study  KdV solitons propagating in the cold quark-gluon plasma (cQGP), in a sense that concerns its spatial profile configuration in the Fourier space. The CE is here proposed to derive the optimal width of the solitonic pulse for which the information stored in its spatial shape are most compressed into its momentum modes.

This paper is organized as follows: Sect. II is devoted to present the model for the cQGP and the associated equation of state, reviewing the energy and pressure densities as functions of the baryon number density and the gluon mass, in the QCD setup. In Sect. III, the KdV equation is derived for baryonic density perturbations in the cQGP. The configurational entropy is then defined, discussed  and calculated for the KdV soliton, as a function of solitary wave pulse width.  Sect. IV is devoted to point out the conclusions, discussions and outlook toward useful generalizations. 

\section{The equation of state in the QGP}

The equation of state for a cQGP model consists of a mean-field approximation of QCD  \cite{celenza,fn4,fn6}.  
It can be derived from a QCD Lagrangian density \cite{Fogaca:2011pk}, 
\begin{equation}
\!\!\!\!\!\mathcal{L}\! =\!\frac{1}{4}F^{a}_{\!\mu \nu}F_{a}^{\mu \nu} \!+\!\sum_{q}^{N_{f}} \!\bar{\psi}^{q}_j\!\left[g\gamma^{\mu}T^{a}_{jk}G^{a}_{\mu}\!+\!\delta_{jk}\! \left(i\gamma^{\mu}\partial_{\mu} \!-\!m_{q}\right)\right]\!\psi^{q}_{k}, 
\end{equation}
where $
F^{a \mu \nu} = \partial^{[\mu}G^{a \nu]}+ gf^{abc}G^{b \mu}G^{c\nu}$. 
The indexes $q$, $i$, $j$ run over all quark flavors and colors respectively, $T^{a}_{jk}$ are the SU(3) generators and the $f^{abc}$ denote its structure constants. The mass of a quark with flavor $q$ is denoted by $m_{q}$, whereas the gluonic field, $G^{a\mu}$, can be split into a condensate component and a high-moment one, as \cite{celenza,Fogaca:2011pk} 
\begin{equation}
G^{a\mu} = A^{a\mu} + \alpha^{a\mu},
\end{equation}
where the $A^{a\mu}$ denote  the soft low-momentum components of the gluon field, which condensate and are related with non-perturbative long-range processes. Besides, the hard-momentum components, $\alpha^{a\mu}$, are dominant in short-distances,  perturbative, processes. Such a splitting requires a specification of a length scale, that must lie in the range $\Lambda_{\rm QCD} \lesssim E \lesssim 1$  GeV. In other words, the $A^{a\mu}$ represent the soft modes populating the vacuum, whereas $\alpha^{a\mu}$ are the modes for which the running coupling is small.   

In order to obtain an effective Lagrangian for the condensate, the mean-field approximation is considered, which means that the hard-modes are coordinate-dependent functions \cite{Fogaca:2011pk,fn4,fn6}, 
$\alpha^{a}_{\mu}(\vec{x},t) = \alpha^{a}_{0}(\vec{x},t) \delta_{\mu 0},$ 
whereas the low-momentum modes are  occupied and coordinate independent. At this point, Refs. \cite{Fogaca:2011pk,fn4,fn6} are followed, wherein the hard component is considered coordinate dependent. This difference in the usual approach, one that assumes $\alpha^{a\mu}$ to be independent of coordinates, leads to an equation of state for the cQGP that allows a solitonic equation for the baryon density perturbation. With such assumptions and replacing the product of the soft modes in the Lagrangian by their expectation values in the cQGP \cite{Fogaca:2011pk,fn4,fn6}, one can obtain the effective Lagrangian density
\begin{eqnarray}
\mathcal{L} &=& -b\phi_{0}^{4} + \frac{m_{g}^{2}}{2}\alpha_{0}^{a}\alpha_{0}^{a} + \bar{\psi}_{i}\left(i\delta_{ij}\gamma^{\mu}\partial_{\mu}+ g\gamma^{0}T^{a}_{ij}\alpha_{0}^{a}\right)\psi_{j}\nonumber\\&& \qquad- \frac{1}{2}\alpha_{0}^ {a}\left({\nabla}^{2}\alpha_{0}^{a}\right).
\end{eqnarray} 
where the first term on the right hand side is related with the dimension four condensate $\langle F^{2}\rangle$, with $b = \frac{9}{4(34)}$, and $\phi_{0}$ is an order parameter driving the condensate. The  hard gluon dynamical mass, $m_{g}$, is  obtained from its interaction with the low-momentum gluons $A^{a\mu}$ and are related with the dimension two gluon condensate $\langle A^{2}\rangle$. The equation of state for the energy density, $\varepsilon= \left\langle T_{00} \right\rangle$, and the pressure density, $p= \left\langle T_{ii} \right\rangle$, can be now  obtained from the effective Lagrangian. 
 From the energy-momentum tensor given by 
\begin{eqnarray}
T^{\mu\nu} &=& \frac{\partial\mathcal{L}}{\partial\left(\partial_{\mu}\eta_{i}\right)}\left(\partial^{\nu} \eta_{i}\right)-  g^{\mu\nu}\mathcal{L} - \left[\partial_{\beta}\frac{\partial \mathcal{L}}{\partial\left(\partial_{\mu}\partial_{\beta}\eta_{i}\right)}\right]\left(\partial^{\nu}\eta_{i}\right) \nonumber \\&& -\frac{\partial \mathcal{L}}{\partial \left(\partial_{\mu}\partial_{\beta}\eta_{i}\right)}\left(\partial_{\beta}\partial^{\nu}\eta_{i}\right).
\end{eqnarray}
the energy density can be written as 
\begin{eqnarray}
\varepsilon &=& \frac{1}{2}\alpha_{0}^{a}\left(\vec{\nabla}^{2}\alpha^{a}_{0}\right) - \frac{m_{g}^{2}}{2}\alpha^{a}_{0}\alpha^{a}_{0} + b\phi_{0}^{4} - g\rho^{a}\alpha_{0}^{a}  \nonumber \\&& + 3\frac{\gamma_{Q}}{2\pi^{2}}\int_{0}^{k_{F}}dk k^{3}.
\end{eqnarray}
where $\rho^{a}$ is the color charge density, $\rho = \psi^{\dagger}\psi $ is the quark number density, $\gamma_{Q}$ is the quark degeneracy factor $\gamma_{Q} = 2 \; ({\rm spin}) \times 3\;({\rm flavor})$, $g$ is a small coupling constant of the hard modes, and $k_{F}$ is the Fermi momentum defined by $\rho = \frac{\gamma_{Q}}{2\pi^{2}}k_{F}^{3}$. The pre-factor 3 on the momentum integral term arises from the sum over all quark colors. It is possible to write the hard modes $\alpha_{0}^{a}$ in terms of $\rho$ and $\rho^{a}$ as \cite{Fogaca:2011pk}
\begin{equation}
\alpha_{0}^{a} = -\frac{g}{m_{g}^{2}}\rho^{a} - \frac{g}{m_{g}^{4}}\vec{\nabla}^{2}\rho^{2}
\end{equation}
and relates the quark color density $\rho^{a}$ and the quark number density $\rho$, by \cite{fn6} $\rho^{a}\rho^{a} = 3\rho^{2}$. With these identities and performing the momemtum integral, the energy density reads 
%

\begin{eqnarray}
\varepsilon &=& \frac{27g^{2}}{2m_{g}^{2}}\left[\rho_{B}^{2} + \frac{\rho_{B}}{m_{g}^{2}}\partial^{2}_x\rho_{B} +\frac{\rho_{B}}{m_{g}^{4}}\partial^ {4}_x\rho_{B} + \frac{1}{m_{g}^{6}}\partial^{2}_x\rho_{B}\partial^ {4}_x\rho_{B}\right]\nonumber\\&&\qquad+ b\phi_{0}^{4} + \frac{3\gamma_{Q}k_{F}^{4}}{8\pi^{2}},
\end{eqnarray}
where $\rho_{B} = \frac{1}{2}\rho$ is the baryon number density, as pointed out the ref. \cite{Fogaca:2011pk}. Similar calculations leads to the expression for pressure density,
\begin{eqnarray}
p &=& \frac{9g^{2}}{m_{g}^{2}}\!\left[\frac{3}{2}\rho_{B}^{2} \!+\!\frac{2g^{2}}{m_{g}^2}\rho_{B}\partial^{2}_x\rho_{B}\!-\!\frac{1}{2m_{g}^{2}}\left(\partial_x \rho_{B}\right)^{2} \right.\nonumber\\&&\left.\,\,+ \frac{1}{2m_{g}^{4}}\left(\rho_{B}\partial^{4}_x\rho_{B}+\left(\partial^{2}_x\rho_{B}\right)^{2}\!\!\!-\partial_x \rho_{B}\partial^{3}_x\rho_{B}\right) \right.\nonumber \\
&&\left.\! -\frac{1}{2m_{g}^{6}}\!\left(\partial^{2}_x\rho_{B}\partial^{4}_x\rho_{B} \!-\! \left(\partial^{3}_x\rho_{B}\right)^{2}\right)\!-\! b\phi_{0}^{4} \!+\! \frac{\gamma_{Q}k_{F}^{4}}{8\pi^{2}}\right].
\end{eqnarray}
In the next section we will carry out a perturbation beyond the linear order on this baryonic density, to obtain a KdV equation, and use the solution to construct the configurational entropy density to explore the it informational content.

\section{The KdV equation and configurational entropy}

Relativistic hydrodynamics is a well-established  paradigm, being useful in many areas of physics. One of its key equations is the  Euler equations, that consist of a system of quasilinear hyperbolic equations that rule adiabatic inviscid flows, and  represent the continuity law and the energy-momentum conservation. Euler equations are particular cases of Navier-Stokes equations, for fluid flows with no viscosity and zero thermal conductivity. The relativistic version of the Euler 
equation  reads \cite{land}
\begin{equation}
\partial_t\vec{v}+(\vec{v} \cdot \vec{\nabla}) \, \vec{v}+{\frac{1}{(\varepsilon + p)\gamma^{2}}}
\bigg({{\nabla} p + \vec{v} \,  \partial_t p}\bigg)=0,
\label{eul}
\end{equation}
where $\vec{v}$ and $\gamma$ are the velocity and the Lorentz factor respectively. Natural units $c=1=\hbar$ shall be adopted hereon. Space and time coordinates will be in unit of [fm]. 
The relativistic version of the continuity equation for the baryon density is $
\partial_{\nu}{j}^{\nu}=0$, reading   \cite{fn4} 
\begin{equation}
\partial_t\rho_{B}+\gamma^{2} \vec{v}  \, 
\rho_{B}\;(\partial_t \vec{v}+ \vec{v} \cdot {\nabla} \vec{v})+{\nabla} \cdot 
(\rho_{B} \, \vec{v})=0,
\label{rhobcons}
\end{equation}
where $\rho_B$ is the baryon density defined on the Section II. In the 1-dimensional  relativistic fluid dynamics, the velocity field is written as 
$\vec{v}=v(x,t) \, \hat{\i}$, where $\hat{\i}$ is the unit vector in the  $x$ direction.  
Eqs. (\ref{eul}) and (\ref{rhobcons}) can be respectively rewritten in the simple form: 
\begin{eqnarray}
\!\!\!\!\!\!\!\!\!\!\!\!\!\!\!\!\partial_t v+v\partial_x v-
{\frac{(v^{2}-1)}{\varepsilon + p}}
(\partial_x p+v\partial_t p)&=&0,
\label{eulu}\\
\!\!\!\!\!\!\!\!\!\!\!\!\!\!v\rho_B(\partial_t v\!+\!v\partial_x v)\!+\!
(1\!-\!v^{2})
(\partial_t \rho_B \!+\! \rho_B\partial_x v\!+\!
v\partial_x \rho_B)&=&0.
\label{rhobconsu}
\end{eqnarray} \\
Now, Eqs. (\ref{eulu}) and (\ref{rhobconsu}) can be merged to derive the KdV equation for the  baryon density perturbations, in the variables 
\begin{equation}
\uprho={\frac{\rho_B}{\rho_{0}}} \hspace{0.2cm}, \hspace{0.5cm} {\rm v}={\frac{v}{c_{s}}}
\label{vadima}
\end{equation}
where $\rho_0$ stands for a central density and $c_s$ denotes the speed of sound. It is worth to note that this $c_s$ is the same usual thermodynamical speed of sound $c_s = \frac{\partial p}{\partial \varepsilon}$, that in this case is hard to be obtained directly from the equations of state (7) and (8), and shall be perturbatively computed in what follows. 
Besides, the  $\zeta$ and $\tau$ coordinates read 
\cite{Fowler}:
\begin{equation}
\zeta=\frac{\sqrt\sigma}{R}(x-{c_{s}}t)
\hspace{0.2cm}, \hspace{0.5cm} 
\tau=\sqrt{\sigma^3}{\frac{{c_{s}}}{R}} t
\label{streta}       
\end{equation} 
where $R$ is a size scale of the cQGP and 
$\sigma$ denotes a parameter ruling the expansion of Eqs.   (\ref{vadima})  $
\uprho=\sum_{k=0}^\infty\sigma^k \rho_{k}$ and ${\rm v}=\sum_{j=0}^\infty\sigma^j v_{j},$ for $\rho_0=1$ and  $v_0=1$. 
Eqs.  (\ref{eulu}) and (\ref{rhobconsu}) were written in the $\zeta-\tau$ space  in Ref. \cite{Fogaca:2011pk}, up to $\sigma^2$.  The notation 
\begin{subequations}
\begin{eqnarray}\upgamma&=&{\frac{27g^{2}\rho_{0}^{2}}{{m_g^2}}}\\
\uptau&=&(\pi\rho_{0}^2)^{2/3}\end{eqnarray}\end{subequations}
 is adopted \cite{Fogaca:2011pk}. Eqs. 
(\ref{eulu}) and (\ref{rhobconsu}) yield 
\begin{widetext}
\begin{eqnarray}
&&\sigma\left[ -(\upgamma c_{s}
+3\uptau\,c_{s})
\partial_\zeta v_{1}+(\upgamma
+\uptau)
\partial_\zeta\rho_{1}\right]
+\sigma^{2}\left[(\upgamma+\uptau)\partial_\zeta\rho_{2}
-c_{s}(\upgamma 
+3\uptau)\partial_\zeta v_{2}
+c_{s}(\upgamma
+3\uptau)\partial_\tau v_{1}
+v_{1}\partial_\zeta v_{1}\right.\nonumber\\
&&\left.\qquad+
\upgamma\rho_{1}\partial_\zeta \rho_{1}
+\uptau\,{\frac{\rho_{1}}{3}}\partial_\zeta \rho_{1}
-2c_{s}[\upgamma 
+2\uptau\,c_{s}]\rho_{1}\partial_\zeta v_{1}
-c_s[\upgamma
+\uptau]v_{1}\partial_\zeta \rho_{1}
+{\frac{3\upgamma}{2R^{2}}}
\partial^{3}_\zeta \rho_{1}\right] =0
\label{eulerexp}
\end{eqnarray}
\end{widetext}
and
\begin{eqnarray}
&&\sigma(\partial_\zeta v_{1} 
- \partial_\zeta \rho_{1})+
\sigma^{2} \left(\partial_\zeta v_{2}
-\partial_\zeta\rho_{2}+\partial_\tau\rho_{1}
+\rho_{1}\partial_\zeta v_{1}\right.\nonumber\\&&\left.\qquad\qquad\qquad\qquad\quad+v_{1}\partial_\zeta \rho_{1}
-c_{s}v_{1}\partial_\zeta v_{1}\right) =0, 
\label{contexp}
\end{eqnarray}
respectively. The first term in Eq. (\ref{contexp})  yields  $\rho_{1}=v_{1}$, implying that 
$c_{s}=\frac{\upgamma
+\uptau}{\upgamma+3\uptau}.
$ 
These results can be replaced into the terms proportional to $\sigma^{2}$,  yielding the KdV equation \cite{Fogaca:2011pk}:
\begin{eqnarray}
&&\!\!\!\!\!\!\!\!\!\!\!\!\!\!\frac{\upgamma}{2R^{2}m_g^2}
\partial^{3}_\zeta \rho_{1}+\bigg(2-c_{s}(1+2\upgamma
+2\uptau)+\frac{\uptau}{3}\bigg)
\rho_{1}\partial_\zeta \rho_{1}\nonumber\\&&\qquad\qquad\qquad\qquad\qquad\qquad+A\partial_\tau\rho_{1}
=0,  
\label{kdv33}
\end{eqnarray}
where $A =\upgamma
+\uptau\,$. 
Returning to the spacetime coordinates yields  \cite{Fogaca:2011pk} 
\begin{equation}
\partial_t \uprho_{1}+c_{s}\partial_x \uprho_{1}
+\alpha {c_{s}} \uprho_{1}\partial_x \uprho_{1}
+\beta\partial^{3}_x \uprho_{1}=0,
\label{kdvqcdg}
\end{equation}
where
\begin{subequations}
\begin{eqnarray}
\alpha &=&{1-\frac{c_{s}^2}{2}}
-\frac{1}{A}\left[(2c_{s}^2-1)\frac{\upgamma}{2}-{{\uptau\,}}\bigg(c_{s}^2
-{\frac{1}{6}}\bigg)\right]\\
\beta &=&\frac{3\upgamma c_s}{m_{g}^{2}A}.
\label{beta}
\end{eqnarray}
\end{subequations}
Eq. (\ref{kdvqcdg}) is the so called KdV equation \cite{Fogaca:2011pk},  whose  solutions are spatially localized pulses with finite energy which propagate with a very small distortion of the characteristic shape of the solitary wave solution.
 
The KdV equation (\ref{kdvqcdg}) has an analytical soliton solution given by
\begin{equation}
\uprho_{1}(x,t)={\frac{3(u-c_{s})}{\alpha c_s}} \sec\!{\rm h}^{2}
\left(\lambda^{-1}(x-ut)\right),
\label{sech}
\end{equation}
where $u$ is an arbitrary supersonic velocity and $\lambda$ is the width of the solitonic pulse,
\begin{equation}
\lambda^2=4\frac{\beta}{u-c_{s}}.
\label{largura}
\end{equation}

Now, the configurational entropy as a logarithmic measure of the information of square-integrable functions shall be employed 
to analyze the KdV solitons arisen in the cQGP. Considering $f(x)$ to be a square-integrable function defined on $\mathbb{R}^{d}$, and its Fourier transform 
\begin{equation}
F(k) = \int_{\mathbb{R}^d} \exp\left(-ik\cdot x\right)f(x) d^{d}x,
\end{equation}  one defines the modal fraction \cite{Gleiser:2012tu}
\begin{equation}\label{modall}
f(k) = \frac{|F(k)|^{2}}{\int |F(k)|^{2}d^{d}k},
\end{equation}
which is the relative weight of a given mode $k$. The configurational entropy of $f(x)$ hence reads \cite{Gleiser:2012tu} 
\begin{equation}
S[f] = - \int_{\infty}^{\infty} \tilde{f}(k)\ln \tilde{f}(k) d^{d}k,
\label{ce}
\end{equation}
where $\tilde{f}(k) = \frac{f(k)}{f_{\rm max}(k)}$ is the normalized fraction that lies in the range [0,1] and also makes the CE to converge.  The informational properties of the definition (\ref{ce}) are more evident when compared with the Shannon's entropy
\begin{equation}
H = - \sum_{i} p_{i} \log p_{i},
\label{shannon}
\end{equation}
where $p_{i}$ are the probabilities of a given symbol of an alphabet $\mathcal{A}$ appears in a string codified in an alphabet $\mathcal{B}$. The amount of information needed to decode such a string depends on the code (a chosen map from $\mathcal{A}$ to $\mathcal{B}$) used. The  more information needed, the less compressed  the information is. Nevertheless, there is a limit in the compression of the information by the code, given an arbitrary  alphabet and any probability distribution. Such a  limit is given by the Shannon's entropy (\ref{shannon}). For the configurational entropy, the spatial shape of a spatially bounded function $
f(x)$ plays the role of the string in the original alphabet, whereas the Fourier transform $F(k)$ is this shape codified in the momentum \textit{alphabet} $\{k\}$. The modal fraction (\ref{modall}) is then the probability distribution associated to a momentum configuration $F(k)$. It measures the contribution of a given momentum configuration to the description of the spatial profile of the function $f(x)$. Following this interpretation, the configurational entropy is a measure of the amount of information needed to described the spatial profile of $f(x)$, by means of its Fourier transform $F(k)$. However, since the Fourier transform is just an isometry of the square-integrable function space (with respect with $L^{2}$ norm), the CE is, in fact, a measure of the informational content of the spatial shape of $f(x)$. 

These ideas are very general, and can be applied to a great variety of problems, provided that one finds some spatially bounded function of the system for which the spatial shape are connected to quantities of interest. It has been accomplished in the past five years pretty much in non-linear and non-equilibrium dynamics \cite{Gleiser:2013mga,Sowinski:2015cfa,Gleiser:2015rwa, Gleiser:2014ipa,Gleiser:2012tu}, but also in others areas  \cite{Casadio:2016aum,Bernardini:2016hvx,Bernardini:2016qit}. In all these works, the interpretation of configurational entropy given above holds. 

In the case of (pure) KdV solitons, by its solitonic nature the shape remains the same as the time evolve, so it is sufficient to consider the solution at an initial time
\begin{equation}
\uprho_{1}(x,t = 0)={\frac{3(u-c_{s})}{\alpha c_s}} \sec\!{\rm h}^{2}
\left(\lambda^{-1}x\right),
\label{sech2}
\end{equation}
and look for extremal points of the configurational entropy with respect to  the width, in order to find whether there is a value of width, consequently a value of velocity, for which the soliton is at  the most or the least information compressibility.    
In order to standardize our results with previous ones \cite{Fogaca:2011pk},  the parameters $g = 0.35$, 
$b\phi_{0}^{4} = -6\times 10^{-4} \,\, \mbox{GeV}^{4}$ and $\rho_{0}=2 \, {\rm fm}^{-3}$ are taken into account, which revealed a  traverse perturbation in the QGP \cite{Fogaca:2011pk}. 



For the model presented in this section, we shall consider hereon $d=1$, corresponding to the 1-dimensional soliton propagation. Since the soliton (\ref{largura}) is already square-integrable, it can be directly used as the configurational entropy density. The KdV solitonic solution given by Eq. (\ref{sech2}) is then taken into account to calculate the associated CE as a function of the soliton width in Eq. (\ref{largura}). Using Eqs. (\ref{modall}) and (\ref{ce}) yields the CE encoded, in the figure in what follows.
\begin{figure}[H]
\centering\includegraphics[width=7.9cm]{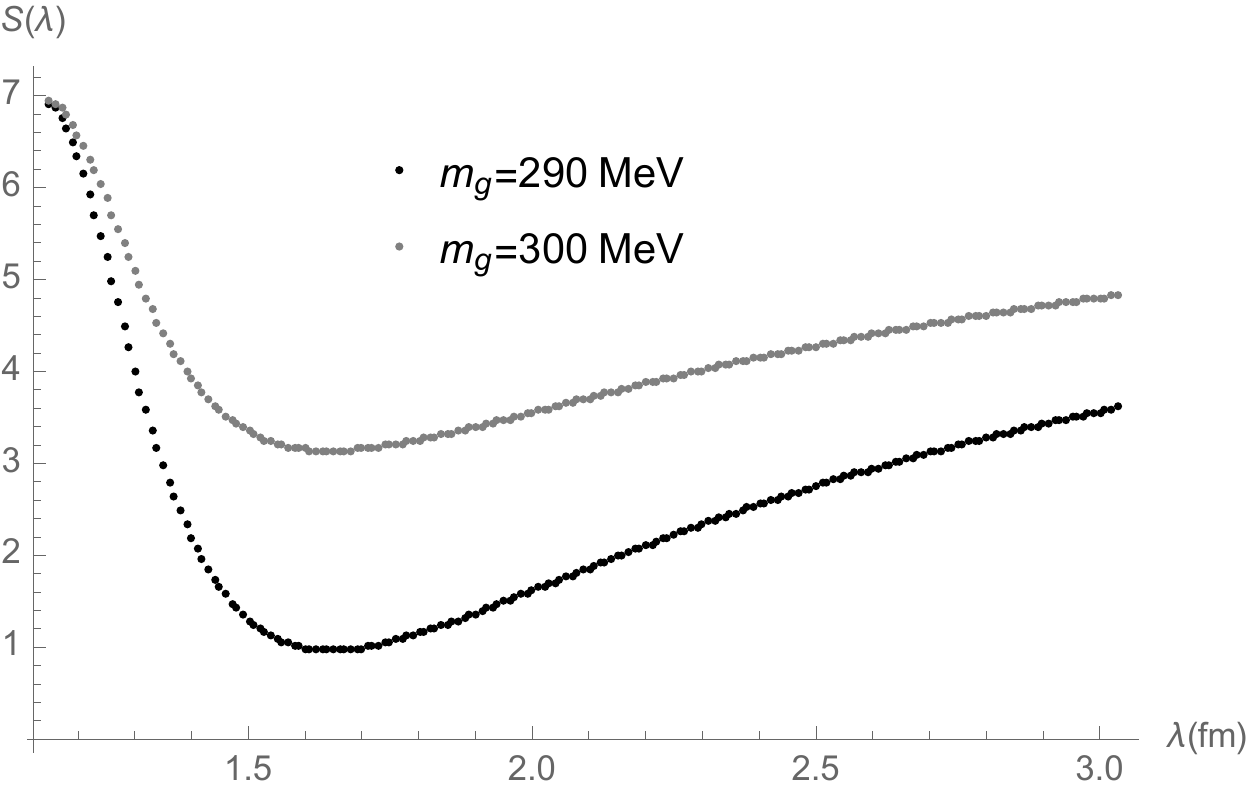}
\caption{Configurational $S(\lambda)$ as a function of the soliton width $\lambda$. For the grey dotted line, the minimum occurs at $\lambda = 1.634$ fm, whereas for the black dotted line the minimum is at $\lambda = 1.613$ fm.}
\label{f1}
\end{figure}
In the above plot, the CE is represented as a  function of the soliton width $\lambda$. For the grey dotted line, the hard gluon dynamical mass is 290 MeV and the most information compressibility configuration 
corresponds to a soliton with width $\lambda = 1.634$ fm. For the hard gluon with dynamical mass 300 MeV, such configuration occurs for a soliton with width $\lambda = 1.613$ fm.

\section{Concluding remarks and outlook}

In this work, the informational content of a KdV solitonic pulse emerged from a perturbation in the baryonic density of a model for a cold quark-gluon plasma, beyond the linear regime are considered. The analysis was based upon the use of the configurational entropy. We showed that theres is a value of the soliton width for which the configurational entropy has a minimum, corresponding to a configuration of maximum compressibility of information in the Fourier modes that describes the spatial shape of the soliton. 

These results show that even for the same solitonic solution of the KdV equation, the variations in the spatial shape induced by different choices of the relevant parameter are non-trivial in an informational sense. The amount of information needed to describe the system is at the most compressed state for neither the minimum width nor the maximum one, as one could have expected. On the other hand, our results show that the configurational entropy can be sensitive to variations of the spatial profile, not only between test functions and exact solutions of a non-linear PDE, as in Ref. \citep{Gleiser:2012first}, or between exact solutions of non-linear PDE with degenerate energy  as in \cite{Gleiser:2017mco,Correa:2014boa}, but even for the same solitary wave solution, with different values of the parameter.

Concerning the cQGP system, we also show that the larger the energy of the condensate (larger hard-gluon dynamical mass), the smaller the width of the informational optimal soliton is. It may   indicate the most likely  KdV solitons to appear as  perturbations of the baryonic density, for each value of the energy, following the conjecture of \cite{Gleiser:2012first}.


An interesting direct extension for this work is the study of a non-zero temperature model of quark-gluon plasma in which solitonic pulses are allowed, in particular a model for the regime of a QGP that is reproduced in the laboratory. Finally, it is interesting to investigate whether the solitonic solutions of modified KdV equations show the same informational behaviour, and if the configurational entropy can detect the emergence of pure solitons as initially mixed states evolving with time.

\acknowledgments
 
The work of  RdR is supported by CNPq (grant No. 303293/2015-2) and  to FAPESP (grant No. 2015/10270-0 and grant No. 2017/18897-8), for partial financial support. The work of A. Goncalves is supported by CAPES.

\end{document}